\documentclass[usenatbib]{mn2e}

\usepackage{gensymb}
\usepackage{graphicx}
\usepackage{amssymb}
\usepackage{epsfig}
\usepackage{natbib}
\usepackage{epsf,palatino,url,graphicx,wrapfig,array,setspace,amssymb,fancyhdr,multirow,lscape,appendix,rotating,amsmath,xcolor}

\def\ion#1#2{#1$\;${\small\rm\@Roman{#2}}\relax}

\title[LXP 27.2 ]{Swift J0513.4-6547 = LXP 27.2 : a new Be/X-ray binary system in the Large Magellanic Cloud}

\author[M.J. Coe et al.,]{M. J.~Coe$^{1}$, M. Finger$^{2}$, E.S. Bartlett$^{3}$, \& A. Udalski$^{4}$ \\
\\
$^{1}$ Physics and Astronomy, University of Southampton, SO17
1BJ, UK. \\
$^{2}$ Universities Space Research Association, Huntsville, AL 35805, USA \\
$^{3}$ Astrophysics, Cosmology and Gravity Centre (ACGC), Astronomy Department, University of Cape Town, Private Bag X3, Rondebosch 7701, South Africa \\
$^{4}$ Warsaw University Observatory, Aleje Ujazdowskie 4, 00-478 Warsaw, Poland \\
}

\begin{document}

\date{Accepted 2014 December 2.  Received 2014 December 2; in original form 2014 November 5}
\maketitle

\label{firstpage}

\begin{abstract}

{An exceptionally bright new X-ray source in the Large Magellanic Cloud was discovered by the {\it Swift/BAT} telescope on MJD 54923 (2 April 2009), and shown to have a pulse period of 27s using follow-up observations by {\it RXTE/PCA} \citep{Krimm09}. We report here on detailed timing observations taken over the following weeks using {\it Fermi/GBM} which reveal an excellent orbital solution and indicate that the source flux peaked at $\sim10^{38}\text{ erg~s}^{-1}$. In addition, we report on follow-up optical observations (spectroscopic and photometric) which permit a classification of the mass donor star as B1Ve, and furthermore reveal a strong optical modulation at a period consistent with the binary period found from the {\it Fermi/GBM} data - 27.4 d. The dynamical mass estimate for the Be star is not in agreement with that expected for a B1V star - a mismatch similar to this has been previously reported for the few other Magellanic Cloud stars that also have dynamical mass estimates. In addition, the neutron star magnetic field determination from the X-ray data ($\sim10^{13}$ G) adds further evidence to the possibility that many such stars in High Mass X-ray Binary systems may have magnetic fields greater than previously expected ($\sim10^{12}$ G).

}

\end{abstract}


\section{Introduction and background}

The Be/X-ray binary systems represent the largest sub-class of all High Mass X-ray Binaries (HMXB).  A survey of the literature reveals that of the $\sim$240 HMXBs known in our Galaxy and the Magellanic Clouds \citep{Liu05,Liu06}, $\ge$50\% fall within this class of binary. In fact, in recent years a substantial population of HMXBs has emerged in the Small Magellanic Cloud (SMC), comparable in number to the Galactic population, though unlike the Galactic population, all except one of the SMC HMXBs are Be star systems.  In these systems the orbit of the Be star and the compact object, presumably a neutron star, is generally wide and eccentric.  X-ray outbursts are normally associated with the passage of the neutron star close to the circumstellar disk \citep{Okazaki01} and generally are classified as Types I or II \citep{Stella86}). The Type I outbursts occur periodically at the time of the periastron passage of the neutron star, whereas Type II outbursts are much more extensive and occur when the circumstellar material expands to fill most, or all of the orbit. General reviews of such HMXB systems may be found in \citet{Reig11}, \citet{Negueruela98}, \citet{Corbet09} and \citet{Coe00,Coe09}

In this paper we present the analysis of X-ray and optical data from the recently discovered transient system Swift J0513.4-6547 in the Large Magellanic Cloud (LMC). The source was initially detected by the {\it Swift} observatory and a pulse period of 27.2s identified \citep{Krimm09}. Following the convention established by \citet{Coe05} of naming pulsating X-ray binary systems in the Magellanic Cloud by their pulse period we will refer to this source as LXP 27.2 throughout this work.

  \citet{Krimm09} associated the X-ray source with the source 2MASS 05132826-6547187. This counterpart was subsequently identified as the R$\sim$15.5 star 59.5431.442 in the MACHO database, but no optical period was detected in those data \citep{Schmidtke09}. Detailed follow-up observations by the GROND optical telescope \citep{Greiner09} established the optical-IR magnitudes of the counterpart and suggested that there was evidence for an optical/NIR brightening over previous archival measurements. Such a brightening is often associated with an X-ray outburst from a Be/X-ray binary system. \citet{Finger09} subsequently reported the detection of the same pulse period in data from the {\it Fermi/GBM} telescope over many days.

\begin{figure}
\includegraphics[angle=0,width=80mm]{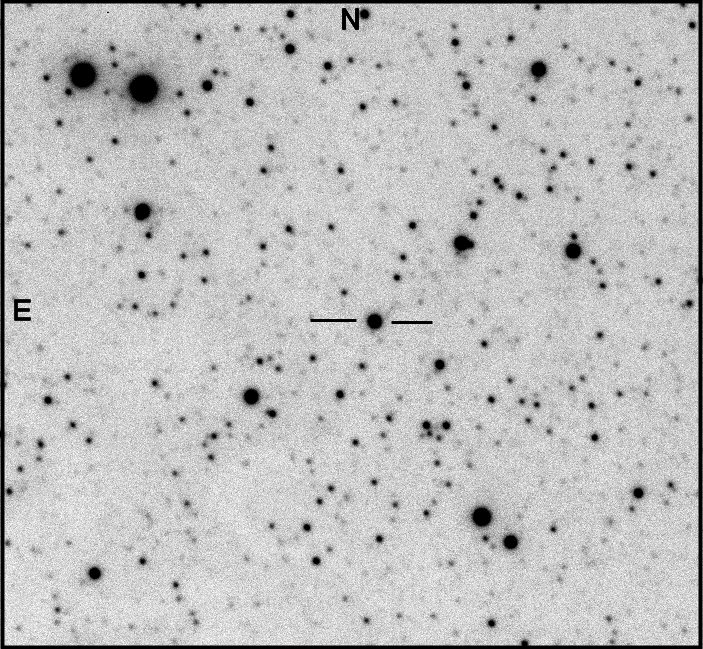}
\caption{ \emph{I}-band image of the 4 x 4 arcmin field around LXP 27.2 taken with the Faulkes Telescope. The optical counterpart is indicated. }
\label{fig:fti}
\end{figure}

The {\it Swift/XRT} telescope detected LXP 27.2 on several occasions and those data can be used to established the best position of the X-ray source as
RA(J2000) = $05^h 13^m 28^s$, Dec(J2000) =$-65^\circ 47' 20''$, with an estimated uncertainty of 1.9 arcsec radius \citep{Krimm09}. This position is consistent with that of a V$\sim$15 star identifiable in several existing catalogues. Figure~\ref{fig:fti} shows an \emph{I}-band image of LXP 27.2.

The source was re-detected by {\it XMM-Newton} and {\it Swift} in Aug and Sep 2014 respectively, both occasions falling close to the time of optical maximum and hence the outbursts were proposed to be evidence of on-going Type I outbursts from this system \citep{Sturm14}.

\section{OGLE and MACHO data}

Optical data from the Optical Gravitational Lensing Experiment Phase IV (OGLE; \citealt{Udalski97,Udalski03}) were used to investigate the long-term behaviour of the optical counterpart in the LXP 27.2 system. The source is identified as LMC 506.16.16 within the OGLE  IV phase of the project.

\begin{figure}
\vspace{-20pt}
\includegraphics[angle=0,width=90mm]{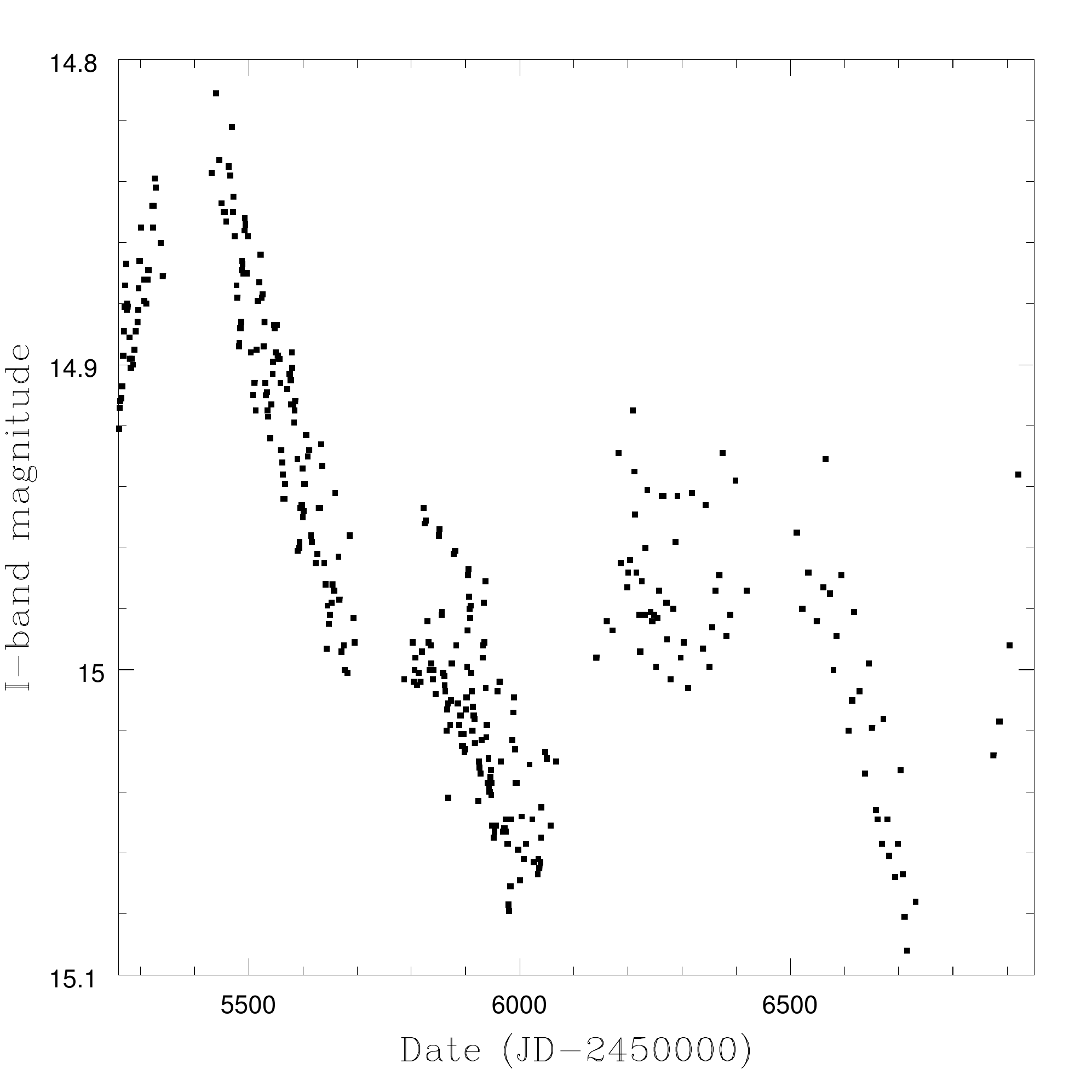}
\caption{ All the OGLE \emph{I}-band data from Phase IV of the project. }
\label{fig:all_ogle}
\end{figure}

Shown in Figure~\ref{fig:all_ogle} are the OGLE IV data covering 4.5 years (MJD 55260 - 56710) - this period starts after the discovery X-ray outburst (MJD 54890 - 54950). It is immediately apparent from the figure that the source exhibits regular optical outbursts of $\sim$0.05 magnitudes in the I band.

\begin{figure}
\includegraphics[angle=0,width=90mm]{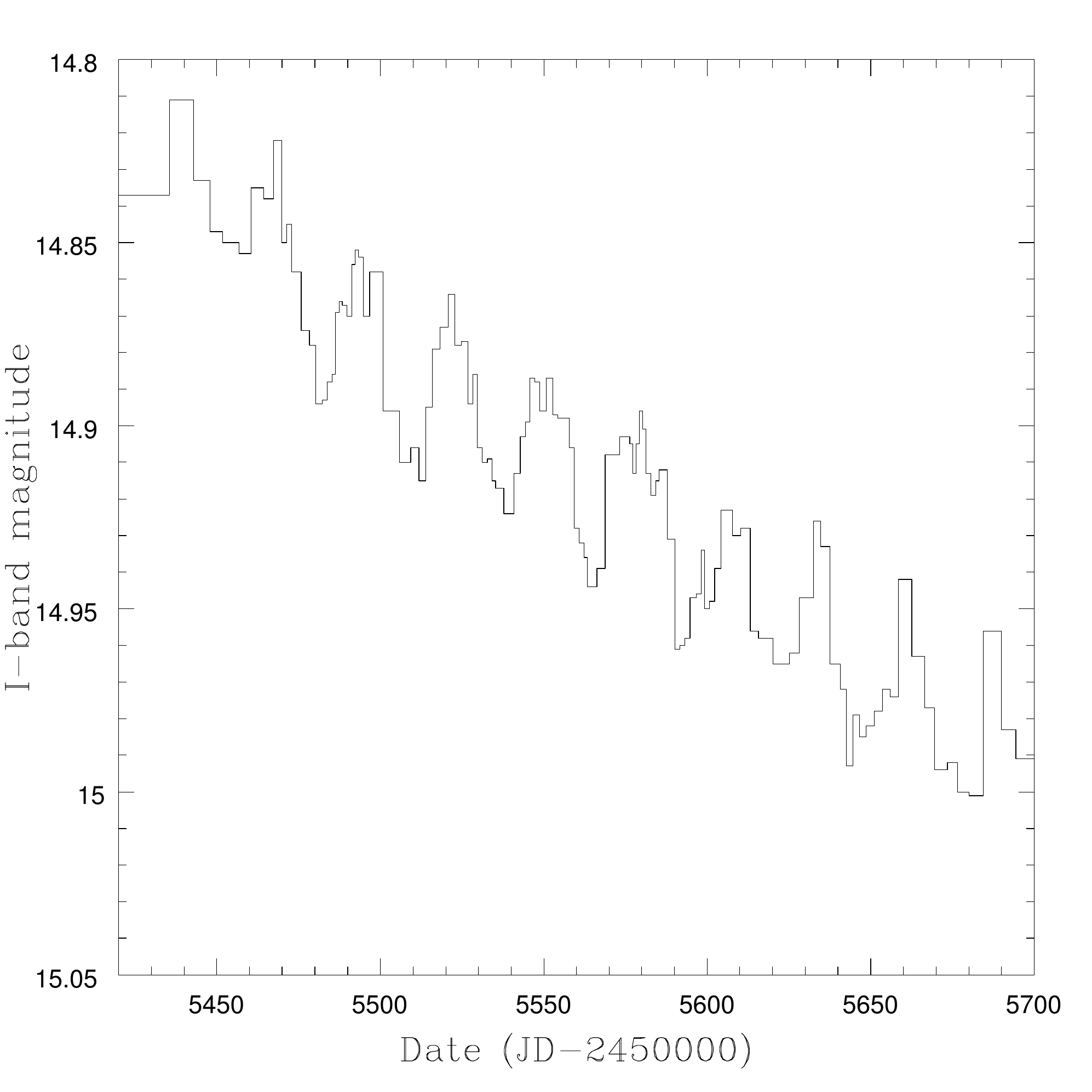}
\caption{ OGLE \emph{I}-band data from the first complete year of observations.
}
\label{fig:ogle1}
\end{figure}

\begin{figure}
\includegraphics[angle=0,width=90mm]{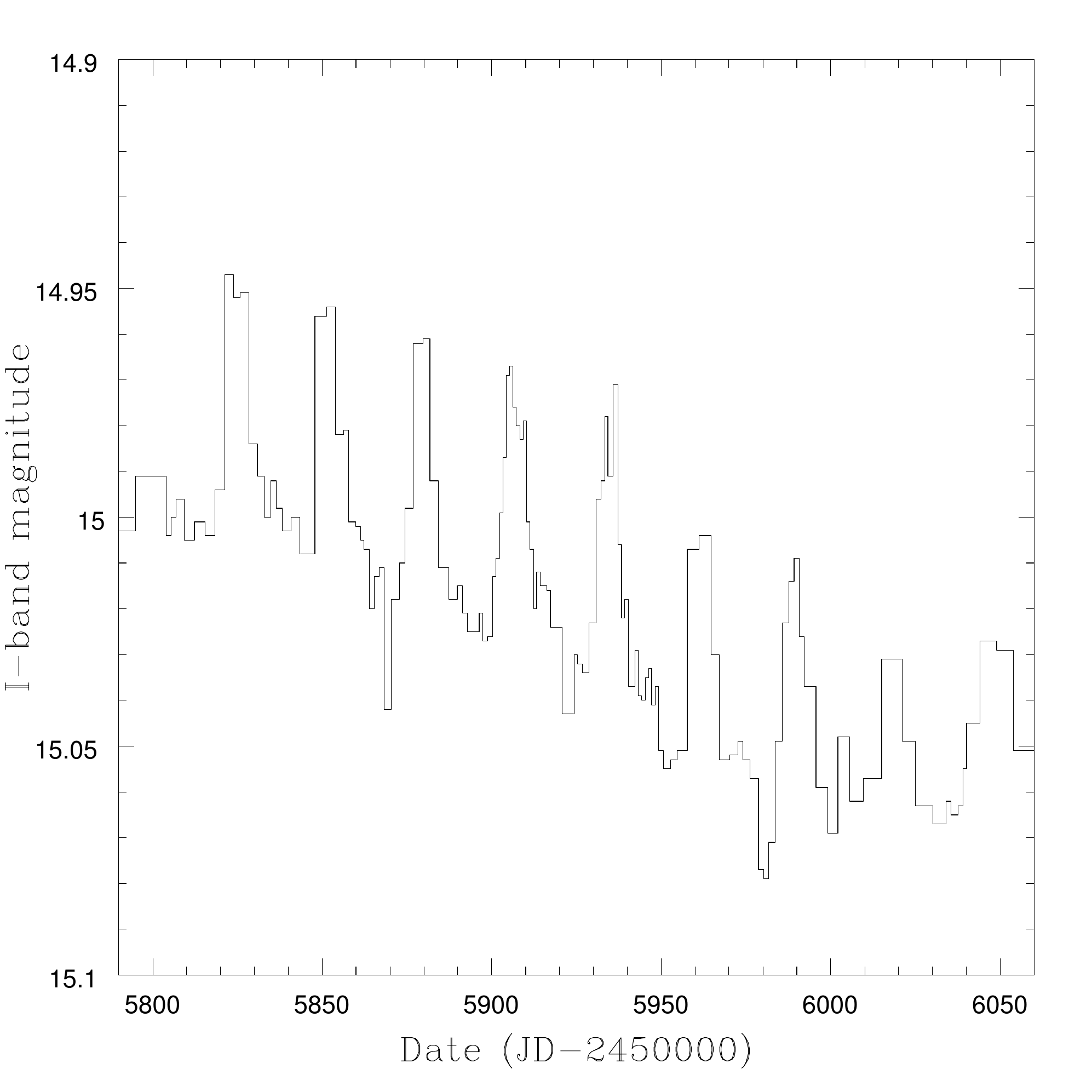}
\caption{ OGLE \emph{I}-band data from the second complete year of observations.
}
\label{fig:ogle2}
\end{figure}

Shown in Figure~\ref{fig:ogle1} are the OGLE IV data covering just the first complete year of observations. From this figure the regular modulation is very clear and reveals a broad modulation with a Full Width Half Maximum (FWHM) of $\sim$12 d. However, shown in Figure~\ref{fig:ogle2} are the OGLE IV data covering the second complete year. There is a strong suggestion that the outbursts are becoming sharper in phase over this year with a FWHM closer to $\sim$9 d.

\begin{figure}
\includegraphics[angle=0,width=90mm]{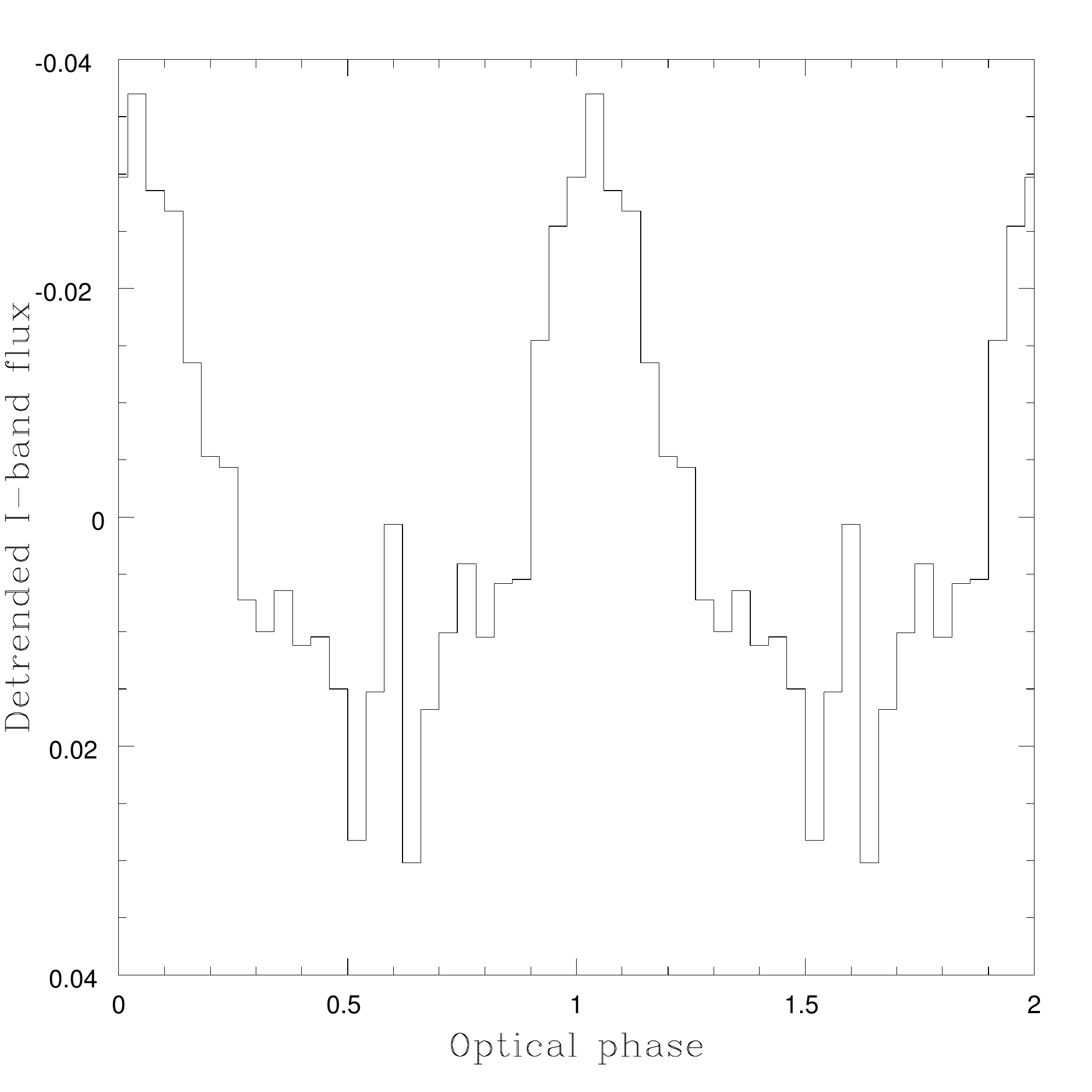}
\caption{ Detrended  \emph{I}-band data averaged over all epochs and folded at the period and ephemeris given in Equation~\ref{eq:ogle}.
}
\label{fig:oglefold}
\end{figure}

A Lomb-Scargle analysis of the whole OGLE  data set after a simple polynomial detrending reveals a strong peak at a period of 27.405d. The averaged, detrended pulse profile is shown in Figure~\ref{fig:oglefold}. Based on the entire data set, we can set an ephemeris for the peak of the optical outbursts to be:

\begin{equation}
\label{eq:ogle}
T_{opt} = (55274.3 + n(27.405\pm0.008)) MJD
\end{equation}

The source was also detected in the much earlier MACHO survey over the period MJD 49127 - 51531 and identified in their catalogue as the LMC star 59.5431.442. Though \citet{Schmidtke09} reported no obvious period in those data, a reanalysis of just the red MACHO data reveals a dominant peak in the Lomb-Scargle power spectrum at 26.6$\pm$0.8d. This is just consistent with the more accurate OGLE period determination.

\begin{figure*}
\vspace{-200pt}
\includegraphics[scale=0.6,angle=0,bb= 0 10 1000 1000]{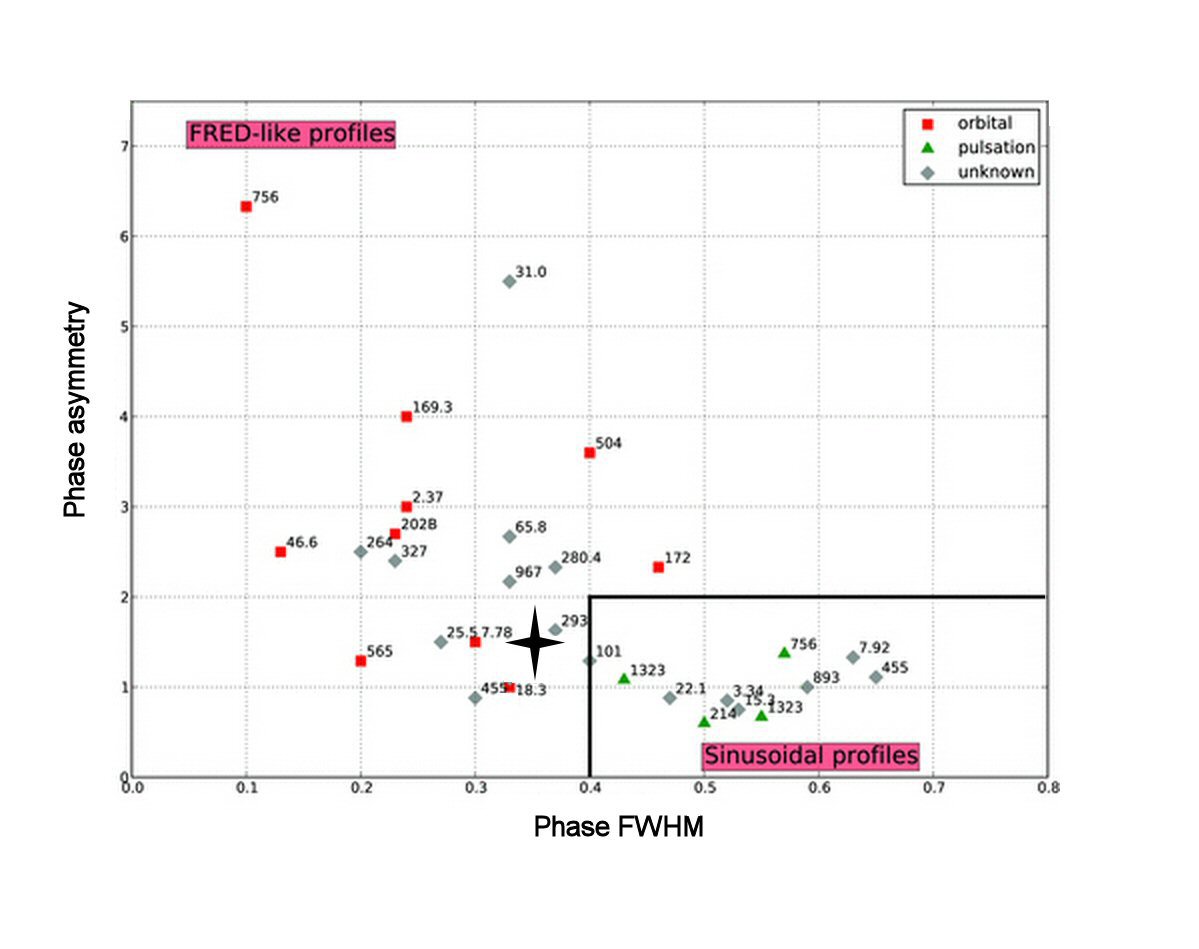}
\caption{ The positions occupied by SMC sources which exhibit optical modulation from the B star companion. The plot shows the folded modulation profile characteristics - taken from \citet{Bird12}. The location of the LMC source LXP 27.2, assuming the same technique applies, is indicated by the star symbol.
}
\label{fig:bird}
\end{figure*}

We can apply the technique discussed in \citet{Bird12} for interpreting such folded light curves. We find that we have a average FWHM for the outburst peak of 0.35, and a value for phase asymmetry of 1.5. This places LXP 27.2 just outside of the zone occupied by non-radial pulsating sources (or period aliases thereof) - see Figure~\ref{fig:bird}. Hence we assume that the 27.405d modulation represents the binary period of the system. It is worth noting that the FWHM value of 0.35 is an average, and the values for each year vary from as high as $\sim$0.5 in Figure~\ref{fig:ogle1} to $\sim$0.25 in Figure~\ref{fig:ogle2}.

\section{SAAO Observations}

The optical data were taken with the 1.9~m Radcliffe telescope at the South African Astronomical Observatory (SAAO) on the night of 2014 September 25 (MJD 56925) with a total observing time of 3000~s split between two observations. A 300 lines mm$^{-1}$ reflection grating was used, blazed at 4600\AA{} with a slit width of 1.8\arcsec along with the SITe CCD. This resulted in a useful wavelength range of $\lambda\lambda3900-6800$~\AA{}.
The resolution of the spectrum is $\sim2.4$ pixels as determined from the FWHM value of Gaussian profiles fit to the arc lines in the comparison spectra, which corresponds to $\sim$5.5\AA{}. The signal-to-noise ratio (SNR) of the spectrum ranges from \textgreater80 at $\sim$55\AA{} to around 30 at $\sim$6500, with the SNR of $\sim$60 over the $\lambda\lambda4400-5000$~\AA{} interval.

The data were reduced using the standard packages available in the Image Reduction and Analysis Facility (\textsf{IRAF}). Wavelength calibration was implemented using comparison spectra of Copper and Argon lamps taken immediately before and after the observation with the same instrument configuration. The spectra were combined, normalised and a redshift correction applied, corresponding to a recession velocity of 280 km s$^{-1}$ \citep{Richter87}.

\begin{figure*}
\centering
 \includegraphics[width=1.0\textwidth]{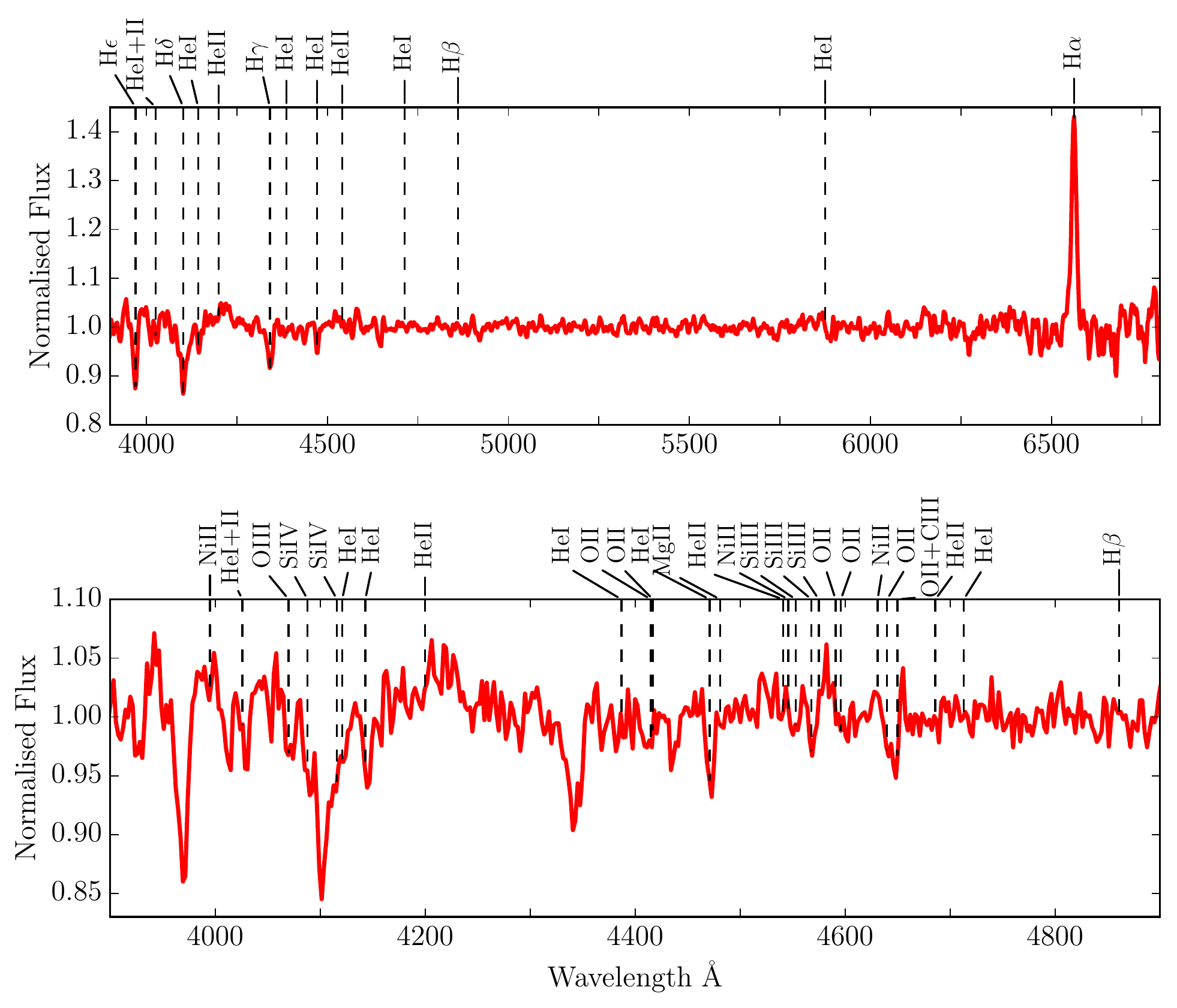}
\caption{Spectrum of LXP27.7 taken with the Radcliffe 1.9~m telescope at SAAO on 2014-09-25. The top panel shows the full $\lambda\lambda3900-6800$\AA{} wavelength range with the prominent Balmer series and Helium lines marked. The bottom panel shows the $\lambda\lambda3900-4800$\AA{} subset of the data along with the weaker metal lines. The spectrum has been normalised and redshift corrected by -280~km~s$^{-1}$.}\label{fig:blue}
\end{figure*}

The top panel of Figure~\ref{fig:blue} shows the full spectrum of LXP 27.2, smoothed with a boxcar average with width 3 pixels, with the prominent Balmer series and Helium lines marked. The bottom panel shows the spectrum, with no smoothing applied, in the $\lambda\lambda3900-4900$\AA{} wavelength range, used for classification.

The spectrum of LXP 27.2 is dominated by the rotationally broadened hydrogen Balmer series. These lines originate in the 'decretion' disc around the Be star which is replenished by the star's rapid rotation. This disc is the source of the emission lines that lead to the 'e' designation. This is well demonstrated by the H$\alpha~\lambda6253$ and H$\beta~\lambda4861$ lines, with H$\alpha$ in emission with an equivalent width of -8.3$\pm1.2$\AA{}. The H$\beta~\lambda4861$ line appears non-existent, the result of an emission line superimposed on an absorption feature.

The low metallicity environment of the Magellanic Clouds makes spectrally classifying early type stars difficult: The metal lines required for classification using the traditional Morgan-Keenan (MK; \citealt{MKK1943}) system are either much weaker or are not present. As such, the optical counterpart of LXP 27.2 was classified using the system developed by \citet{Lennon97} for B-type supergiants in the SMC, and implemented for the SMC and LMC
by \citet{Evans06}. This system has been normalised to the MK system such that the classification criteria follows the same trends in line strengths.

The spectrum does not show any evidence for the He\textsc{ii} $\lambda\lambda4200, 4541  \text{ or}\ 4686$ $\AA{}$ lines above the noise level of the data, suggesting a spectral type later than B0. The Si\textsc{iv} $\lambda4088$ $\AA{}$ line is clearly present suggesting a spectral type B1 or later. The relative strength of this line compared with those of the O\textsc{ii} spectrum confirm this classification. The absence (or presence) of the Si\textsc{iv} $\lambda4116$ $\AA{}$ is difficult to ascertain due to the line's proximity with the Doppler broadened H$\delta$ line, however the presence of Si\textsc{iii} and absence of the Mg\textsc{ii}$\lambda4481$ line constrains the spectral type to earlier than B2.5.

The luminosity class of the counterpart was determined using the ratio of He\textsc{i} $\lambda4121$/He\textsc{i} $\lambda4143$, which is anti correlated with the Si\textsc{iv}$\lambda4553\text{/He}\textsc{i}\lambda4387$ ratio recommended for classification by \citet{Walborn90}. The former decreases with increasing luminosity class (i.e. from I-V; decreasing luminosity) the latter increases with increasing luminosity class. This line ratio suggests a luminosity class V.  The signal-to-noise and resolution of the spectrum make it difficult to draw any firm conclusions based on this spectrum alone, however we note that this spectral and luminosity class are in good agreement with those derived below from the  photometric data, as such we classify the optical counterpart of LXP 27.2 as a B1V star.

\section{Faulkes Telescope observations}

The optical counterpart to LXP 27.2 was observed remotely at an airmass of 1.45 using the southern Faulkes Telescope \citep{Brown13} on MJD 56576 (11 Oct 2013) Data were collected under excellent observing conditions and the field containing the source was imaged in all the standard Johnson filters (\emph{U}, \emph{B}, \emph{V}, \emph{R} \& \emph{I}). Data on the standard star E267 were also taken to calibrate the frames. The resulting magnitudes are presented in Table~\ref{tab:ft}.

\begin{table}
\centering
  \caption{Optical magnitudes for the counterpart to LXP 27.2 determined in this work from the Faulkes Telescope (FT). Other magnitudes come from UCAC4 \citep{Zacharias13}, USNO \citep{Monet03} and OGLE IV (this work).}
  \label{tab:ft}
  \begin{tabular}{ccccc}
  \hline
  Band& FT & UCAC4 & USNO & OGLE  \\
  \hline
  \emph{U} & 14.18 &&& \\
  \emph{B} & 15.17 & 14.92 & 15.3 & \\
  \emph{V} & 14.96 & 14.95 & & \\
  \emph{R} & 14.90 & 15.45 &  15.5 & \\
  \emph{I} & 14.86 &&& 14.8 - 15.1 \\
  \hline
  \end{tabular}
  \end{table}

In order to estimate the spectral class from these magnitudes we need to adjust for the reddening to the LMC of E(\emph{B--V})=0.07 \citep{Schlafly11} and for a distance modulus of 18.5 \citep{Walker12}. To avoid any potential contamination from the circumstellar disc surrounding the OB star, just the \emph{U} \& \emph{B} magnitudes were used for the following estimate of the spectral class. Correcting the observed magnitudes results in a (\emph{U--B}) colour of -1.03 which corresponds to a B0V star.

\begin{figure}
\centering
\hspace{-20pt}
\includegraphics[angle=-90,width=0.5\textwidth]{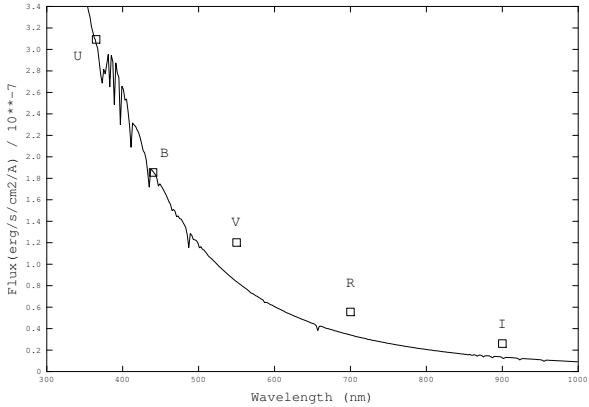}
\caption{Observed optical magnitudes for the counterpart to LXP 27.2 corrected for extinction and distance to the LMC. The resulting magnitudes have been converted to flux and compared with a Kurucz model for a B1V star (continuous line). The model is normalised in the \emph{U} band.
}
\label{fig:model}
\end{figure}

Assuming this is the correct spectral classification, we can see the extent to which the other bands are modified by emission from the circumstellar disk - see Figure~\ref{fig:model}. By fitting a Kurucz model for a B0V star (T=30000K, log (g)=4.) to the \emph{U} \& \emph{B} points it is immediately apparent that there is a considerable excess present redward of the \emph{B} band. It is, of course, possible that even the \emph{B} band is slightly contaminated and this would modify the final spectral classification, so we adjust our estimate of the spectral class to B0-B1V in agreement with the spectral classification.

\section{X-ray observations}

\subsection{{\it Swift} observations}

\begin{figure}
\includegraphics[angle=90,width=80mm]{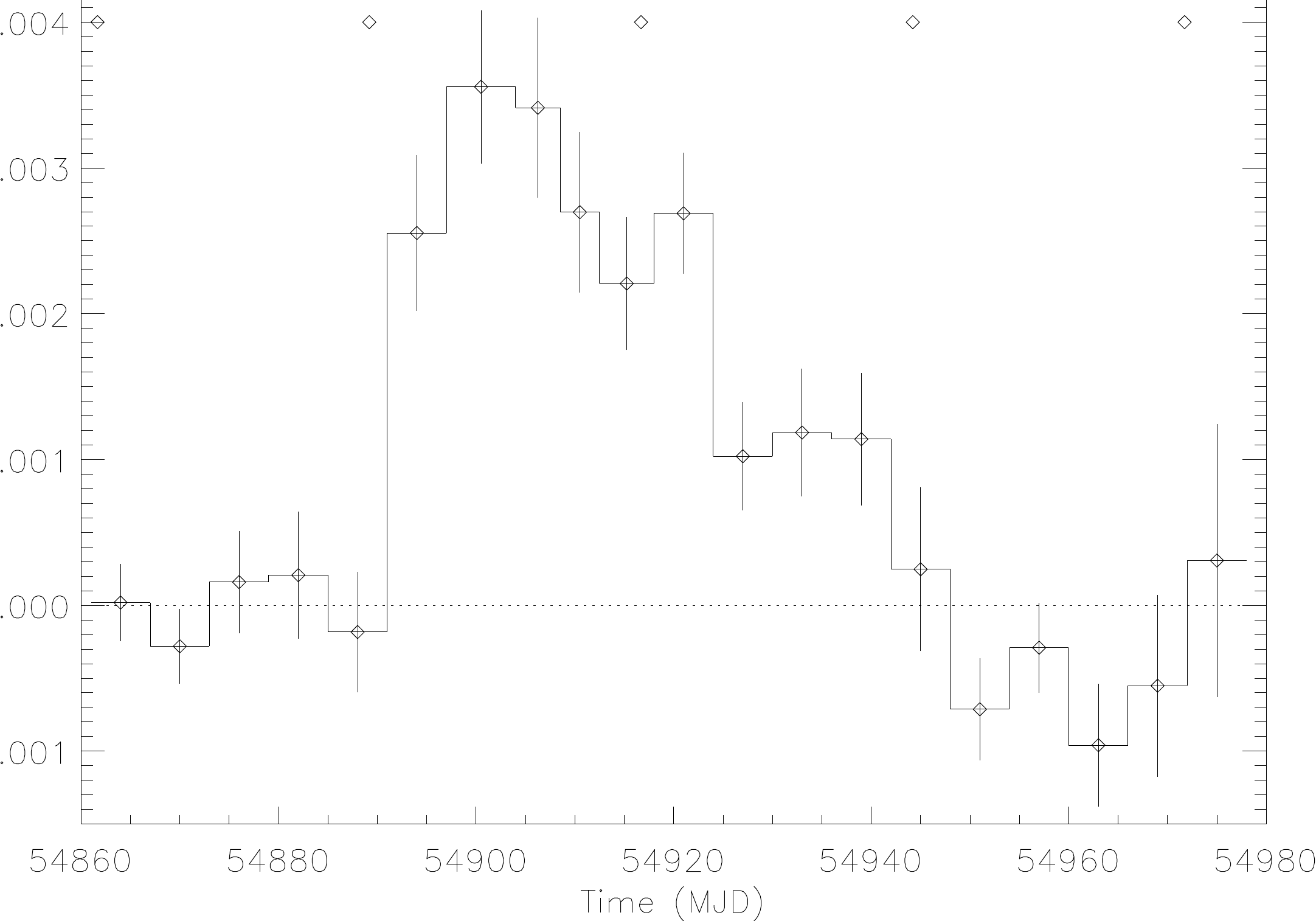}
\caption{ {\it Swift/BAT} count rates averaged over 6 day intervals. The diamonds indicate the phases of the optical peaks according to the ephemeris given in Equation~\ref{eq:ogle}.
}
\label{fig:bat}
\end{figure}

Shown in Figure~\ref{fig:bat} are the {\it Swift/BAT} count rates throughout the outburst. As we do not have simultaneous optical photometry for this period we have indicated the phases of the optical peaks seen subsequently in the OGLE data and given in Equation~\ref{eq:ogle}. From this we can see that if the optical peak occurs at periastron, then the X-ray outburst began close to this phase - perhaps not surprising as this is the point at which the neutron star would be closest to the circumstellar disk.

\begin{figure}
\includegraphics[angle=-90,width=80mm,clip=true,trim=1.5cm 0cm 0cm 0cm]{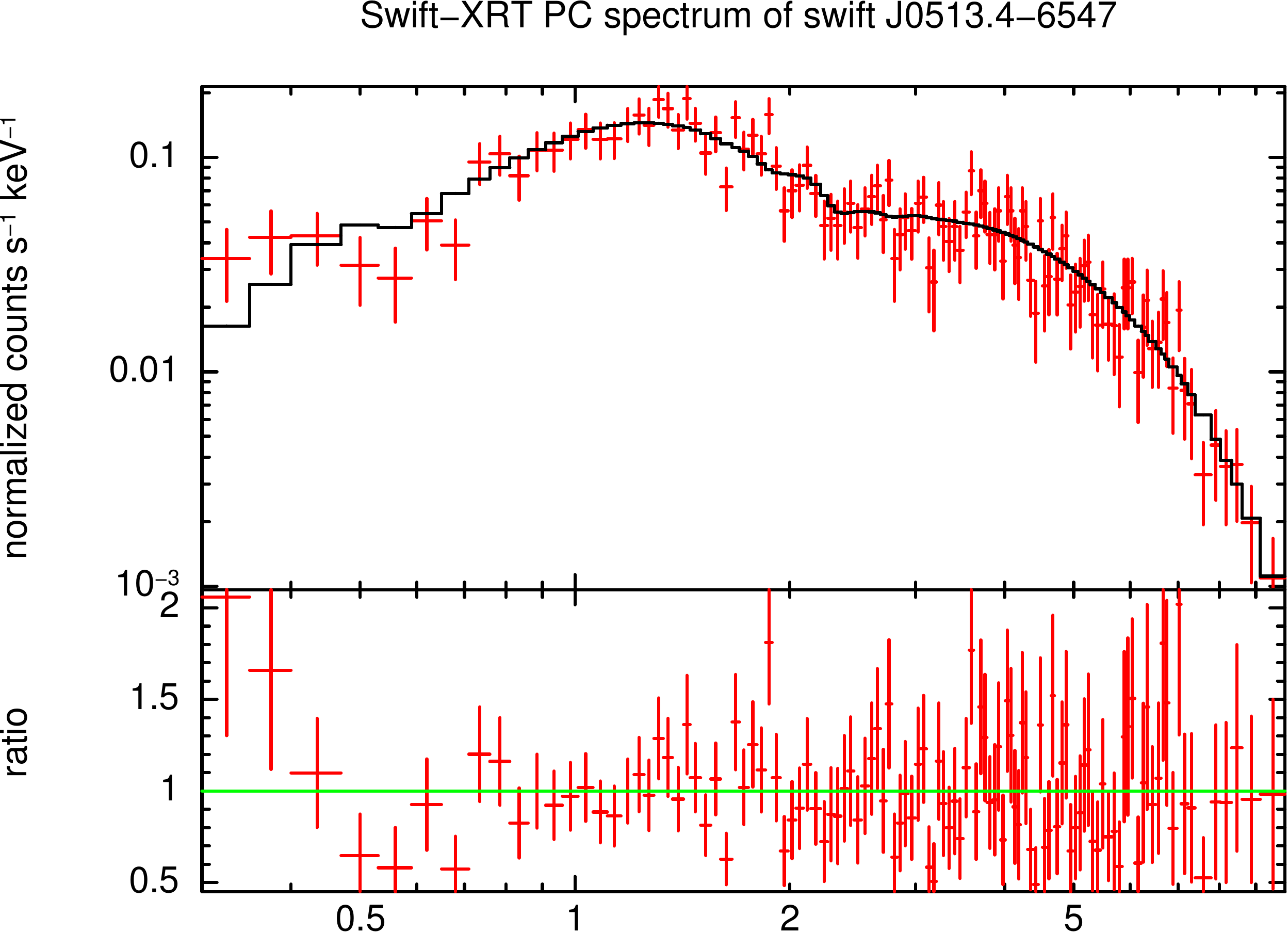}
\vspace{10pt}
\caption{ Swift/PC time averaged spectrum from the three 2009 observations.
}
\label{fig:pc}
\end{figure}

The source was also detected in the {\it Swift/XRT} instrument on several occasions in 2009 \& 2014. The three 2009 observations had a total exposure time of 4.6 ks. Shown in Figure~\ref{fig:pc} is the time averaged spectrum of the 2009 data. The parameters of the simple power law fit are presented in Table~\ref{tab:pc}. The photon index is typical of that seen in Be/X-ray systems and the observed average flux corresponds to a luminosity of $1.3\times10^{37}$ \text{erg~s}$^{-1}$ at the distance to the LMC of 50 kpc \citep{Pietrzynski13}.

\begin{table}
\centering
  \caption{Best fit Swift/PC X-ray absorbed power-law model parameters from the 2009 observations. }
  \label{tab:pc}
  \begin{tabular}{cc}
  \hline
  Spectral parameter& Value  \\
  \hline
$N_{H}$ & $(1.3\pm0.5) \times 10^{21}$ \\
&\\
Photon index & $0.92\pm0.09$ \\
&\\
Observed Flux  & $(4.4\pm0.3) \times 10^{-11}$ \\
(0.3-10) keV & $\text{erg~cm}^{-2}~\text{s}^{-1}$ \\
  \hline
  \end{tabular}
  \end{table}

\subsection{{\it Fermi} observations}

For pulse timing analyses we have used the 12 - 50 keV data of the 12 {\it Fermi/Gamma Ray Burst Monitor (Fermi/GBM)}  Na\textsc{i} detectors. Pulsations from LXP 27.2 were detected from MJD 54891.0 to 54925.5. The data analysis is discussed in \citet{Camero10}.

\begin{figure}
\includegraphics[angle=0,width=80mm]{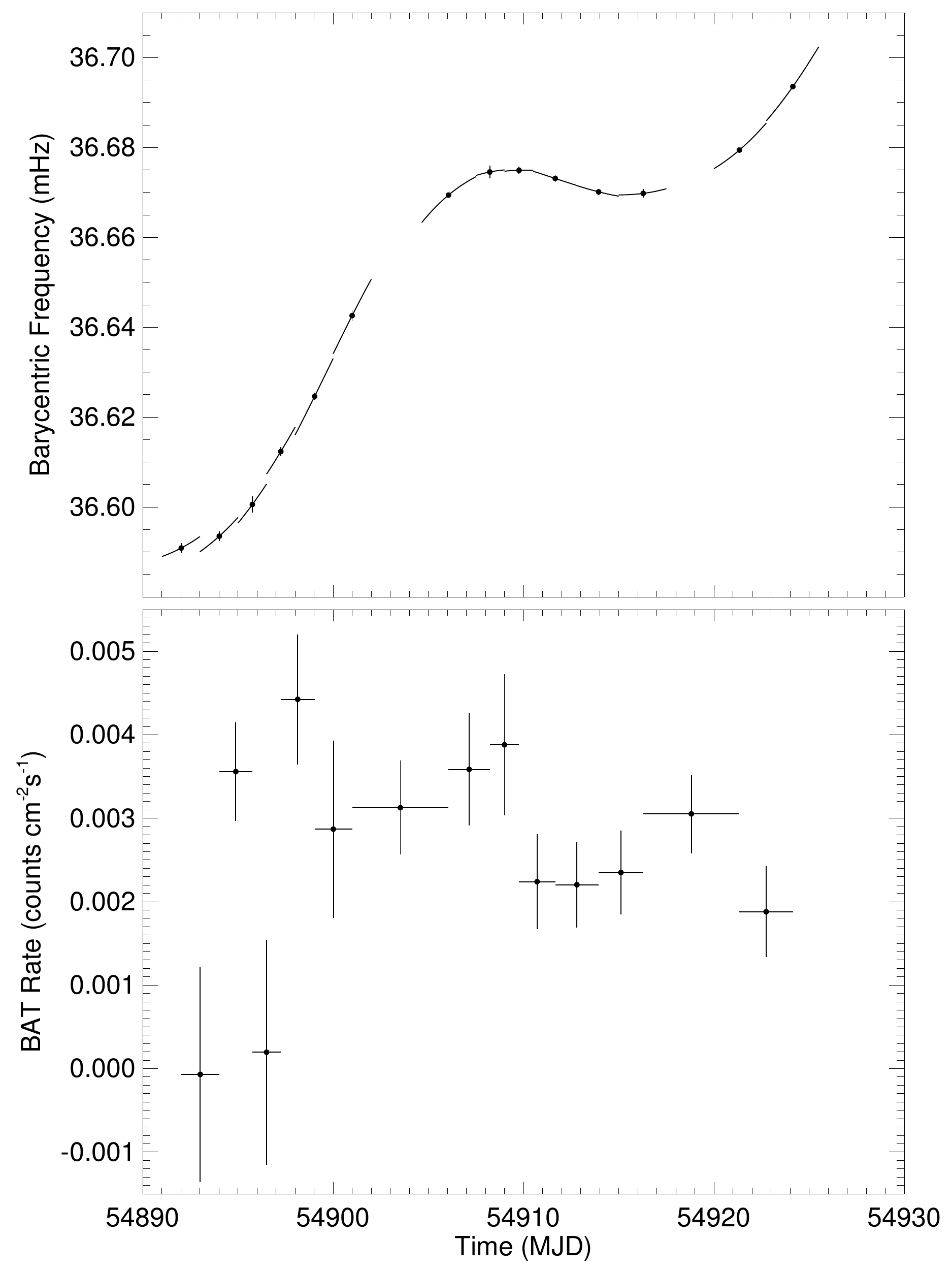}
\caption{The top panel shows the barycentric frequencies determined from {\it Fermi/GBM} measurements. The lower panel shows the {\it Swift/BAT} count rate. }
\label{fig:fermi1}
\end{figure}

\citet{Finger09} made blind searches for pulsation from LXP 27.2 using these {\it Fermi/GBM} data. These searches included both trial frequencies near the 36.66 mHz (27.28 s period) Rossi X-ray Timing Explorer ({\it RXTE}) measurement \citep{Krimm09}, and trial frequency rates in the range expected for accreting pulsars near Eddington luminosity. The source was detected in five broad time intervals between 2009 March 1 and March 28 (MJD 54891.0 - 54918.0). These showed the barycentric frequency rate rising to  $1.0^{-10} \ {\rm Hz \ s}^{-1}$, then dropping to  $8^{-12} {\rm Hz \ s}^{-1}$, suggesting a combination of accretion induced torque and an orbital signature.

Assuming that the optical period represents the orbital period, the orbital signature can be separated from the spin-up trend. At high accretion rates an accretion disk is expected to be present, and the spin-up rate, $\dot{\nu}$, will be proportional to $L^{6/7}$ where $L$ is the luminosity \citep{Ghosh79}. Using the {\it Swift/BAT} count rate in the 15-50 keV range as a surrogate for the luminosity, the spin-up rate can be modelled as $\dot{\nu} = mC^{6/7}$ where $C$ is the BAT rate and $m$ a constant.

Initially we used an ad-hoc procedure to estimate the spin-up trend and orbital elements from the \citeauthor*{Finger09} measurements. First $C^{6/7}$ was integrated between two times separated by the orbital period, then $m$ was set to the change in the barycentric frequency between these two times divided by the integral. The spin-up trend was then calculated from the BAT rates, and subtracted from these frequencies, thereby leaving a constant frequency plus an estimate of the orbital signature. From this, initial determinations of the orbital elements were made.

Next we made new searches for pulsations in frequency and frequency rate using the pulsar time system, as given by the initial orbital elements. To provide more time resolution a set of shorter search intervals were used. This was possible because fewer search trials were required for each interval since narrower search ranges, based on the approximate orbital parameters and spin-up rate model, could be used. Detections where made in 8 intervals between MJD 54891.0 and 54925.5. Barycentric frequencies were computed and a joint $\chi^2$ fit of the barycentric frequencies and the BAT rates was made with a model of the form:
\begin{eqnarray}
   F_i &=& \nu_i(1-\beta_i) \\
   X_i &=&  m^{-1} \, \dot{\nu_i}
 \end{eqnarray}
\begin{itemize}
\item $F_i$ is the measured barycentric frequency at time $t_i$ which is the frequency epoch of search interval $i$. Each epoch is chosen as the mean exposure-weighted observation time.
    \item $X_i$ is the average value of $C^{6/7}$ between $t_i$ and $t_{i+1}$.
    \item The model parameter ${\nu_i}$ is the orbitally corrected frequency at time $t_i$.
        \item $\beta_i$ is the orbital redshift factor at time $t_i$ which is a function of the orbital elements.
            \item $\dot{\nu_i}$ is $(\nu_{i+1}-\nu_i)/(t_{i+1}-t_{i})$.

\end{itemize}

            Using the resulting orbital elements new pulse searches were made, and the fit updated.

Finally we made a model of the orbitally corrected frequency history using the previous orbital elements and measurements, selected a set of shorter intervals, and made new pulse searches. For these shorter intervals the frequency rate was set by the model, with only frequencies searched. Detections were made in 14 intervals. Barycentric frequencies were computed for these intervals, and fit with the BAT rates from the model discussed above. The upper panel of Figure~\ref{fig:fermi1} shows the barycentric frequencies with the slope shown for each interval based on the the model. The lower panel shows the average BAT rate between neighboring frequency epochs.

\begin{table}
\centering
  \caption{Best fit binary model parameters. $T_{\pi/2}$ is the epoch when the mean orbital longitude equals 90$^o$. $a_x \sin(i)$ is the semi-major axis. $g$ = $e\sin(\omega)$ and h = $e\cos(\omega)$ with $e$ the eccentricity, and $\omega$ is the longitude of periastron. m is discussed in the text and $P_{orbit}$ is the fixed orbital period.}
  \label{tab:fermi}
  \begin{tabular}{cc}
\hline
Orbital parameter& Value \\
\hline
 $T_{\pi/2}$ & MJD 54899.02 $\pm$ 0.27 \\
 $a_x \sin(i)$ & 191 $\pm$ 13 lt-s \\
 $g$ & 0.06 $\pm$ 0.04 \\
 $h$ & -0.03 $\pm$ 0.04 \\
 $m$ & (5.4$\pm 0.3) 10^{-9}\text{~Hz ~s}^{-1} \text{(cm}^2\text{~s)}^{6/7}$ \\
 $P_{orbit}$ & 27.405 $\pm$ 0.008 days \\
 e & $\le$0.17 \\
 \hline

  \end{tabular}
  \end{table}

With the orbital period fixed to 27.405 days the $\chi^2$ was 9.80 with 8 degrees of freedom.
The resulting best fit parameters are presented in Table~\ref{tab:fermi}.

The significance of the eccentricity is low. If $g$ and $h$ are fixed to 0 the $\chi^2$ increases to 12.73 with 10 degrees of freedom. If the orbit were circular, then the F test shows a 50\% probability of higher $\chi^2$ difference occurring by chance. Fits for the 2-d array of eccentricity perimeters sets a 95\% confidence upper limit to the eccentricity of 0.17.

If all the parameters including the orbital period are fit, the $\chi^2$ drops to 9.67 with 7 degrees of freedom. The orbital period estimate is $P_{orbit}$ = 27.10 $\pm$ 0.83 days, consistent with the optical period.

\subsection{Other reported X-ray observations}

 The source was observed by {\it RXTE} over the period MJD 54935 - 54953 (14 Apr 2009 - 2 May 2009) and the results are presented in \citet{Krimm13}. They confirm the pulse period originally detected by {\it RXTE/PCA} \citep{Krimm09} and present weak evidence for a general spin-down trend occurring after the period of time reported here from the {\it Fermi} data. There is also evidence that the flux was falling rapidly, so the {\it RXTE} observations obviously caught the source in the final stages of its outburst.

Subsequently, \citet{Sturm14} report more recent detections of LXP 27.2 on MJD 56894 (25 Aug 2014) with {\it XMM-Newton}, and with {\it Swift} on MJD 56916 (16 Sep 2014). From Equation~(\ref{eq:ogle}) we can infer that the optical phases of these observations are
0.02 and 0.82 - both close to optical maximum (phase 0.0) and presumably to the periastron passage of the neutron star. Such a recurrence of activity in LXP 27.2 is supported by the last few data points in Figure~\ref{fig:all_ogle} which show a definite re-brightening of the optical counterpart since MJD 56800. If this continues we can expect a return to the source exhibiting significant X-ray flux in the near future.

\section{Discussion}

\subsection{Constraints on the Be Star Mass}

\begin{figure}
\resizebox{3.125in}{!}{\includegraphics{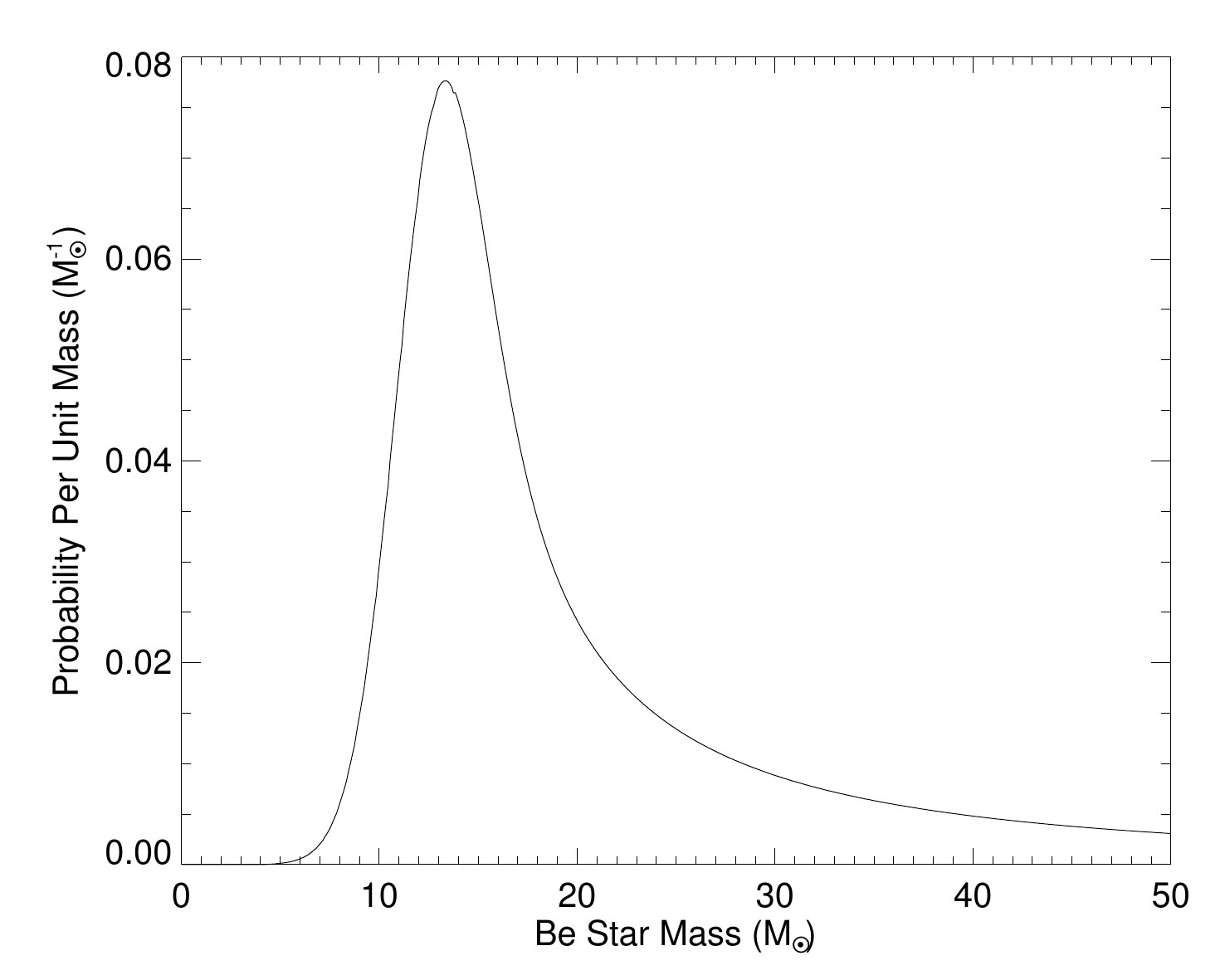}}
\caption{The probability distribution for the companion's mass. The pulsar's mass is assumed to be 1.4 $M_\odot$, and the pole of the binary orbit is assumed to be random. The error on the mass function is treated as gaussian.}
\label{fig:mass}
\end{figure}

The mass function
\begin{equation}
  f(M) = \frac{4\pi^2(a_x\sin i)^3}{G P_{orbit}^2} = \frac{(M_c \sin i)^3}{(M_c+M_x)^2}
\end{equation}
provides a constraint on the companion star's mass. Here $M_c$ is the companion star's mass and $M_x$ is the pulsar's mass. The measured $f(M)$ is 9.9 $\pm$2.0 $M_\odot$. If we assume  that the pole of the binary orbit is randomly oriented and the pulsar's mass is 1.4 $M_\odot$, a probability  distribution for the companion's mass can be constructed. This is shown in Figure~\ref{fig:mass}. The probability peaks at 13.3 $M_\odot$, with 50\% probability in the range of 9.9 - 19.1 $M_\odot$. The  upper tail of the distribution extends to high masses, with 19.5\% probability above 50 $M_\odot$.

The proposed classification discussed above of the companion (B1V) would imply a mass of 16--18$M_\odot$ which lies within the 50\% range, but some way from the peak of the distribution. However, this may once more indicate the mismatch between dynamical determined stellar masses, and that expected based upon the spectral classification for OB stars in the Magellanic Clouds (see also \citealt{Coe14} for another example). Discrepancies of this kind between stellar classification and observed luminosities plus inferred masses \& temperatures, have previously been reported for HMXB companions by Conti (1978) and Kaper (2001).

\subsection{Constraints on neutron star parameters}

The estimated fit parameter $m$ in Table~\ref{tab:fermi} gives the relationship between the spin-up rate $\dot{\nu}$ and $C$

\begin{equation}
\dot{\nu} = (5.4 \pm 0.3)\times10^{-9}\, {\rm Hz\, s^{-1} } \times \left(\frac{C}{cm^{-2} s^{-1}}\right)^{6/7} \label{nudot:bat}
\end{equation}

We convert this to a relationship between spin-up rate, $\dot{\nu}$, and luminosity, $L$, first determining the relationship between luminosity and BAT rate, and then applying Equation \ref{nudot:bat}.
Since there was no broad-band spectral measurement made during the outburst, we compared the time-averaged 5-200 keV spectra of A 0525+26 measured with \emph{INTEGRAL} in February 2011 \citep{Caballero11} to the time-average BAT count rate during the observation and found a ratio of $1.2\times10^{-7}$ erg. We believe the ratio for LXP 27.2 will be within 30\% of this value. For the distance to LXP 27.2 of 50 kpc this corresponds to a luminosity to BAT rate ratio of $3.6\times10^{40}\text{ erg~cm}    ^2$. The peak BAT rate in Figure~\ref{fig:bat} therefore corresponds to a luminosity of $1.3\times10^{38}~\text{erg~s}^{-1}$. Applying the luminosity to BAT rate ratio to Equation~\ref{nudot:bat} we get the relationship
\begin{equation}
\dot{\nu} = 3.5 \times 10^{-11} \, {\rm Hz \,s^{-1}} \times L_{38}^{6/7} \label{nudot:L}
\end{equation}
where $L_{38}$ is the luminosity of units of $10^{38}\text{erg~s}^{-1}$.

Comparing this with theoretical models which gives
\begin{equation}
\dot{\nu}_{-11} = 1.1 \xi^{1/7} M_{1.4}^{10/7}  R_{12}^{2/7}  B_{12}^{2/7} L_{38}^{6/7} \label{nudot:L2}
\end{equation}
where $\dot{\nu}_{-11}$ is the spin up rate in units of 10$^{-11}$ Hz s$^{-1}$, $\xi$ is a factor between 0.5 and 1 that modifies the inner radius of the accretion disk, $R_{12}$ is the radius of the neutron star in unites of 12 km, and $B_{12}$ magnetic field in unit of 10$^{12}$ G. This model can be derived from Equations 2-4 of \citet{Bildsten97}, the assumption of a dipole magnetic field, and treating the moment of inertia as that of a uniform sphere.

The relationships in Equations \ref{nudot:L} and \ref{nudot:L2} can be brought into concordance by changing some of the neutron star parameters from their unit values. However if we do this by only changing the mass, the required mass is above 3{M$_\odot$}, indicating a black hole. If we change only the magnetic field, the cyclotron line energy is above 511 keVF. Bringing the neutron star mass within the limits currently observed to be possible (e.g. \citealt{Demorest10, Antoniadis13}), we find that a 2.0 solar mass neutron star with a 10$^{13}$ G magnetic field is a solution. This result adds to the increasing evidence that magnetic fields exceeding the canonical 10$^{12}$ G may be present in such systems (e.g. \citealt{Eksi14,Klus13,Klus14}). In fact, the value proposed here for LXP 27.2 of 10$^{13}$ G fits very well on the general period versus magnetic field plot for Be/X-ray systems in the Small Magellanic Cloud (see Figure 6 of \citealt{Klus14}).

\section{Conclusions}

LXP 27.2 is an interesting new Be/X-ray binary source in the LMC. Its general behaviour is very similar to other such systems, but the detailed measurements reveal the possibility of an exceptionally strong magnetic field in the neutron star partner. Furthermore this is only the second system in the Magellanic Clouds to have the mass of the OB star determined from dynamical measurements, and like the first one, there seems to be significant discrepancies between the mass estimates from the spectral class and the direct dynamical determination. These results encourage closer looks at the other systems when the opportunities arise.

\section{Acknowledgements}

The OGLE project has received funding from the European Research Council under the European Community's Seventh Framework Programme (FP7/2007-2013)/ERC grant agreement no. 246678 to AU. This work makes use of observations from the LCOGT network, obtained through the Faulkes Telescope Project. ESB is supported by a Claude Leon Foundation Fellowship.

\bsp

\label{lastpage}

\end{document}